\documentclass[conference]{IEEEtran}

  	\usepackage[pdftex]{graphicx}
  	\graphicspath{{../pdf/}{../jpeg/}}
	\DeclareGraphicsExtensions{.pdf,.jpeg,.png}

	\usepackage[cmex10]{amsmath}
	\usepackage{mathabx}
	\usepackage{algorithmic}
	\usepackage{array}
	\usepackage{mdwmath}
	\usepackage{mdwtab}
	\usepackage{eqparbox}
	\usepackage{url}
	\usepackage{cite}
\begin{document}

\title{\LARGE Leveraging Machine Learning Force Fields (MLFFs) to Simulate Large Atomistic Systems for Fidelity Improvement of Superconducting Qubits and Sensors}

% \author{\authorblockN{Leave Author List blank for your IMS2013 Summary (initial) submission.\\ IMS2013 will be rigorously enforcing the new double-blind reviewing requirements.}
% \authorblockA{\authorrefmark{1}Leave Affiliation List blank for your Summary (initial) submission}}

 \author{\IEEEauthorblockN{Søren Smidstrup, Shela Aboud, Ricardo Borges, Anders Blom, Pankaj Aggarwal, Robert Freeman, and Jamil Kawa \\  Synopsys, Inc \\ Sunnyvale, CA 94085 \\Email: {soren, sjaboud, ricardo, andersb, pankaja, ref, jamil}@synopsys.com }
  }

\maketitle

\begin{abstract}
Materials engineering using atomistic modeling is an essential tool for the development of qubits and quantum sensors. Traditional density-functional theory (DFT) does however not adequately capture the complete physics involved, including key aspects and dynamics of superconductivity, surface states, etc. There are also significant challenges regarding the system sizes that can be simulated, not least for thermal properties which are key in quantum-computing applications. The QuantumATK tool combines DFT, based on LCAO basis sets, with non-equilibrium Green’s functions, to compute the characteristics of interfaces between superconductors and insulators, as well as the surface states of topological insulators. Additionally, the software leverages machine-learned force-fields to simulate thermal properties and to generate realistic amorphous geometries in large-scale systems. Finally, the description of superconducting qubits and sensors as two-level systems modeled with a double-well potential requires many-body physics, and this paper demonstrates how electron-electron interaction can be added to the single-particle energy levels from an atomistic tight-binding model to describe a realistic double-quantum dot system. 

\end{abstract}

\IEEEoverridecommandlockouts

\IEEEpeerreviewmaketitle

% ===================
% # I. Introduction #
% ===================

\section{INTRODUCTION}
Paradigm shifts in human society have always been closely linked with radical innovations in materials, from the metallurgical advancements during the iron and bronze ages to the modern computer era founded on silicon. It is therefore logical that to take the next leap into a world where quantum computers can solve the most complex problems yet envisioned, materials engineering will play a crucial role, specifically in the development of the fundamental building blocks underpinning this novel technology, viz. qubits and sensors. And just like the most advanced current traditional semiconductor technology has been enabled and accelerated by computational materials modeling on multiple levels, so will the development of quantum computing rely on robust procedures for simulating both the materials making up the qubits and sensors themselves, and their operation.
The simulation tools needed for these tasks will be required to provide capabilities that go beyond the standard codes used in materials research today. Apart from the obvious fact that the operation of a qubit is an inherently quantum-mechanical process, the materials that make up the qubit also exhibit many specialized features that derive from quantum physics. An obvious example is the superconducting qubit, but alternative technologies built on e.g. topological insulators are also being considered, and spin degrees of freedom are often manipulated and exploited to define qubit states as singlet and triplet levels. At the same time, qubits are relatively large objects compared to single molecules and other systems routinely simulated with quantum-mechanical models, and external or environmental influences from magnetic fields and temperature play crucial roles in these systems, both for creating the effects that make the qubits work, and in the processes that may make them fail. All in all, these factors make materials science for qubit development particularly challenging, but by using recently developed simulation techniques such hurdles can nevertheless be overcome. \\  \\
In this paper we will showcase examples on how atomistic modeling can be used to aid and accelerate the materials development of qubit technology, and discuss the specific simulation tools and methods required for these calculations to be time-efficient and accurate.

% =======================================================
% # II. Impact of traps on large signal characteristics #
% =======================================================

\section{MODELING CHALLENGES FOR COMPLEX, LARGE-SCALE QUANTUM-MECHANICAL SYSTEMS}
Given the observations above, it would be natural to use a continuum quantum-mechanical approach to model qubits, since size effects are less problematic, and fields and temperature can be introduced via model parameters. However, such simulations rely heavily on a detailed database of physical properties for each material to be studied, and two complicating factors make this a daunting or even impossible task. First of all, a primary use case for modeling is to find or optimize novel materials with desired properties, rather than just considering a few well-known options. Second, many of the effects that underpin the qubit operation occur at interfaces or surfaces,  where the continuum description may break down or at least needs some serious extra work to account for such boundary conditions, and moreover many of the systems of interest exhibit size-dependent effects due to quantum confinement etc, which again makes it very difficult to pre-compute all possible material descriptors. \\ 

We therefore need to turn to discrete modeling techniques, where a flexible and efficient choice is offered by density-functional theory (DFT). This first principles or ab initio modeling technique provides complete freedom to study basically any material made up of any combination of atoms from the entire periodic table, without requiring any pre-existing calibration of parametrization. The only input to the calculation is the atomic structure under consideration, i.e. the position and type (chemical element, and perhaps isotope) of every atom in the system. In principle, DFT can be used for both electronic structure calculations and extraction of mechanical/thermal properties, but in practical terms the system sizes set limits on this, at least for dynamical simulations. For truly large-scale electronic structure simulations we may therefore instead turn to semiempirical tight-binding (TB) for less time-consuming calculations, at least provided we can use DFT to parametrize the TB models when needed. \\ 

Moreover, empirical potentials or forcefields can be used for efficient molecular dynamics (MD) simulations of systems with millions of atoms. Here, however, we face a fundamental challenge since accurate parametrizations of such potentials only exist for a small subset of materials, and rarely for the ones of interest for qubit development. Developing a forcefield for a new material has traditionally been considered a very demanding task, but recent developments in optimization techniques based on machine-learning (ML) algorithms have radically changed this situation, and it is now possible to construct a new machine-learned forcefield (MLFF) for a novel material – even complex ones involving interfaces and multiple different chemical elements – with reasonable effort. This is particularly important since it opens up for ways to account for temperature-dependent effects also in DFT calculations through electron-phonon interaction or by simply generating thermodynamic ensembles of atomistic structures and including the temperature effect in a statistical manner. \\

It is now clear that materials engineering of qubits needs a multi-model simulation approach is needed, even when confining the scope to discrete, atomistic simulations. A major challenge is the traditional separation of DFT, semi-empirical and forcefield models into separate codes. To address this issue, the QuantumATK atomistic modeling platform from Synopsys \cite{smidstrup2020} incorporates all mentioned methods in a single, homogeneous framework, which not only provides a seamless transition in model complexity from forcefields to tight-binding to DFT and GW calculations, but also enables many synergetic benefits. A calculation can, e.g., easily use the electronic structure from DFT while the lattice vibrations or phonons are described by an MLFF, in order to compute electron-phonon coupling parameters for
superconducting materials. Primarily using localized orbitals or so-called LCAO basis sets for DFT, QuantumATK offers unprecedented capability to run very accurate large-scale ab initio simulations for systems containing several thousand atoms on modest hardware resources and with reasonable turn-around time. When needed for accuracy, it is also possible to switch plane-wave basis sets. \\

On top of this, QuantumATK also includes a unique Green’s function model, coupled with DFT and tight-binding and using appropriate open boundary condition, to overcome the further challenge of studying electron transport on the quantum-mechanical atomistic level. A variant of this technique can also be used to compute the surface band structure of topological insulators, which conventional DFT with periodic or finite boundary conditions fails to correctly describe. \\

In this paper we will provide several examples of how the QuantumATK platform can be used in each of the three common scenarios where atomistic modeling provides time-to-solution benefits for materials engineering, here specifically applied to qubits: study and explain mechanisms on a fundamental physical level, guide experiments by option down-selection (materials screening), and parametrize higher-level models in a multiscale or hierarchical workflow. We will also, briefly, discuss how to model a quantum-mechanical two-level system – the fundamental operating principle of most qubits – using a discrete tight-binding model. \\ 

\section{UNDERSTANDING THE IMPACT OF VARIABILITY IN SUPERCONDUCTING TUNNEL JUNCTIONS}
For a functioning quantum computer to be viable, it will be necessary to manufacture hundreds if not thousands of identical qubits operating within very stringent variability margins. However, before one can even begin to optimize the process conditions that influence to the functional variability, it is crucial to understand the underlying causes of the variations themselves. This is a prime use case for atomistic modeling, where one can easily play around with a large number of hypothetical structures, without worrying about how to experimentally realize them. The studies are also “embarrassingly parallel” since each structure is an independent calculation that can be run concurrently, provided sufficient hardware resources are available, making such virtual experiments very time efficient. \\ 

As an example, one may consider the prototypical fundamental component of a qubit, viz. a Josephson junction (JJ) formed by an insulator (I) sandwiched between two metal (more precisely, superconductor, S) leads. The insulator layer in this kind of S/I/S system is typically an amorphous oxide, but basic understanding of how the exact structure of this barrier affects the performance and deficiencies of the qubit is severely lacking. Modeling amorphous materials presents its own set of general challenges, and the complexity of the interfaces with the metals only adds to this. \\ 

\begin{figure}[ht!] %!t
\centering
\includegraphics[width=3.5in]{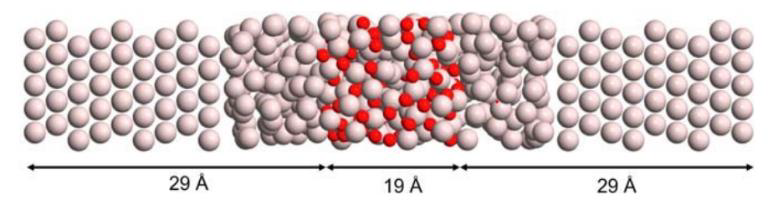}
\caption{S/I/S junction model (Al/AlO$_x$/Al) from Ref.\ \citen{lapham2022}.}
\label{fig:junction_model}
\end{figure}

\begin{figure}[ht!] %!t
\centering
\includegraphics[width=3.5in]{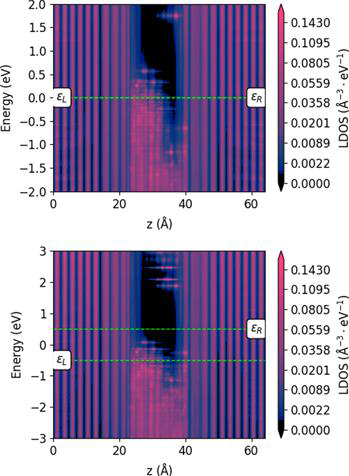}
\caption{PLDOS of AlO$_x$ barriers at zero bias (top) and 1~V bias (bottom)}
\label{fig:barriers}
\end{figure}

Using QuantumATK, it is possible to study such a system using a combination of efficient atomistic electronic structure models that can handle large-scale systems (thousands of atoms) and non-equilibrium Green’s functions (NEGF) to compute the current–voltage characteristics, resistance and critical current of the junction. These calculations should be performed for an ensemble of structures which explore the oxide stoichiometry and density, and the inherent variability that arises due to small differences in the atomic structure across the oxide barrier, as well as the barrier length itself. Such calculations on an Al/AlO$_x$/Al junction (Figs.\ \ref{fig:junction_model} and \ref{fig:barriers}) have shown that the aluminum-oxygen coordination sensitively affects the resistance and critical currents by orders of magnitude. \cite{lapham2022}  \\ 

A crucial first step underpinning these calculations is an efficient algorithm that is able to generate many realistic structures with relevant variability. This typically involves MD simulations in order to amorphize the oxide, and many steps of global geometry optimization to find defect-free realistic interface structures. Such calculation would be far too time-consuming to perform with DFT, but, as mentioned already, in the (common) case no classical forcefield exists for the material of interest. In this situation, QuantumATK allows users to fit a fast and accurate customized MLFF for the material system at hand, a process facilitated by ready-made protocols for amorphization, interface generation, and ML-based fitting/testing of the forcefield. In other cases one may opt to use one of the several built-in general purpose MLFFs that are pre-fitted for almost the entire periodic table, although calculation times are then usually longer and the accuracy noticeably lower.  \\ 

Ultimately, by being able to pinpoint the most important contributors to the variability by this kind of “what-if” modeling, engineers will be able to focus their resources on optimizing the most relevant factors of their device design. \\ 

Some TMDs are topological insulators and can therefore be used as quantum spin Hall insulators (QSHI). Unlike in a regular insulator, the surfaces of a topological insulator can support an electrical current. In the reduced dimensionality of topologically insulating TMDs, the electrons rather travel along a one-dimensional edge, and provided that time-reversal symmetry is not broken, these edge states will be gapless and are protected from backscattering. \cite{jelver2019} This unique property means that a QSHI can protect a qubit from external perturbations and enable robust quantum computation. \\ 

If, however, time-reversal symmetry is broken, the topological edge states may be partially or completely destroyed. Such symmetry breaking is often associated with external magnetic field or intrinsic magnetic impurities. A recent study has showed that time-reversal symmetry may even be broken spontaneously by the presence of magnetism without any impurities, simply because a TMD monolayer can acquire magnetic edge states despite the bulk 2D material being non-magnetic. Thus, one cannot rely on assumptions about a particular TMD based on screening of bulk properties to ascertain its suitability as a QSHI platform for qubits, but a deeper study of its edge states is required. \\ 

\begin{figure}[ht!] %!t
\centering
\includegraphics[width=3.5in]{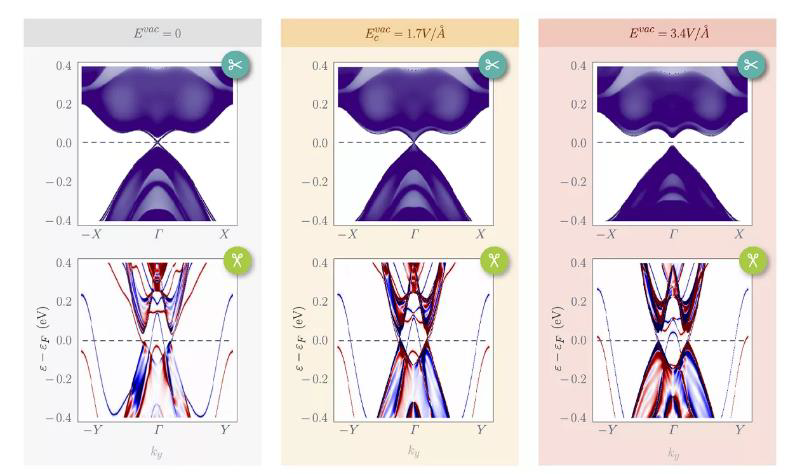}
\caption{Surface band structure of an MoS$_2$ monolayer along the (a) “X” edge, and (b) “m” edge. The top row shows the sum of all spin components, whereas the bottom row shows with red/blue the two separate spin channels}
\label{fig:structure}
\end{figure}

To compute correctly edge states (or more generally surface states) in a material, the standard periodic boundary conditions used in DFT calculations cannot be used, since by necessity such a system will contain a bottom and a top surface in a so-called slab configuration. The artificial confinement induced by this will result in an incorrect, finite gap in the degenerate surface/edge states. Instead, one must employ specialized surface Green’s function techniques coupled with DFT which allow a proper description of a single interface (edge) and its electronic states, as shown in Fig. \ref{fig:structure}. QuantumATK is the only code which has implemented such a model, making the tool a unique and ideal platform for the exploration of not only two-dimensional topologically insulating TMDs but also bulk topological materials such as Bi$_2$Te$_3$ or MgB$_2$. \\ 

The broader class of MB$_2$ diborides (M may be one of several metal ions) moreover exhibits superconductivity, making them particularly interesting for qubit research. In addition to studying the topological surface properties using QuantumATK, Ref.\ \citen{{an2021}} also used DFT to estimate the critical temperature $T_c$ of these materials. The results were in excellent agreement with experiments, thus providing confidence that DFT can be used to screen material candidates which combine a high critical temperature and desired topological properties. Additionally, the paper showed that applying strain can enhance $T_c$, providing novel avenues for realizing high-temperature super-conductivity.  \\

\section{ATOMISTIC MODEL OF QUANTUM DOT TWO-LEVEL SYSTEMS}

\begin{figure}[ht!] %!t
\centering
\includegraphics[width=3.5in]{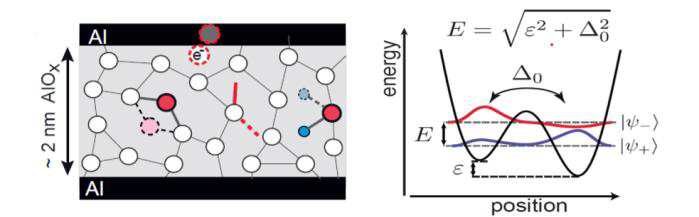}
\caption{Double-well modeling a TLS with a double-well potential, taken from Ref.~\citen{muller2019}.}
\label{fig:double_well}
\end{figure}

As a final example of how discrete simulations can be helpful in qubit design, we consider an atomistic representation of a two-level system (TLS), which can be used as a model for various material defects that couple to a superconducting qubit and reduce its coherence time. This is schematically shown on the right-hand side of Fig. \ref{fig:double_well}. By using a discrete quantum-mechanical description of the system, we can compute the energy level splitting and other properties directly from electronic structure theory, while accounting for material-specific factors for the actual semiconductor used, like non-parabolic electronic bands, finite band width far from the Gamma point, spin-orbit coupling strength, and valley splitting due to strain. \\ 

QuantumATK provides a fully atomistic calculation to describe the interaction between two interacting electrons in a double quantum dot system. Since the system is relatively large (over 1 million atoms, typically), we resort to semi-empirical tight-binding instead of DFT. The interdot gate is modeled as two harmonic potentials wells, which is included in the calculation as an external potential that is added to the regular Hamiltonian. In a more advanced scenario, this potential could equally well be taken from a device level TCAD simulation, and in general have any desired form, with subatomic resolution. \\

The single-particle eigenvalues and eigenstates are evaluated in a narrow energy interval using a non-self-consistent tight-binding model, which is orders of magnitude faster than if we performed a full diagonalization (typically we limit the calculation to less than one percent of the spectrum). These are then used to create corresponding many-body eigenvalues and eigenstates using a configuration interaction (CI) model. The common approach here is to use Slater-Condon rules between Slater determinants to obtain the one and two-particle integrals contributions; notably, the two-particle integrals can be very time-consuming to evaluate, so careful algorithms tuning and parallelization is required for reasonable turn-around times. \\ 

\begin{figure}[ht!] %!t
\centering
\includegraphics[width=3.5in]{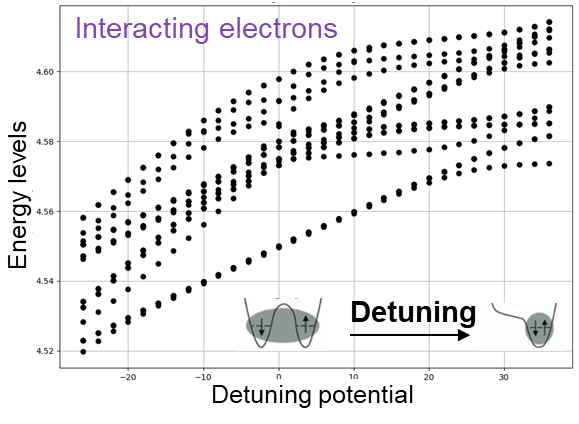}
\caption{Many-body energy levels for interacting electrons in a GaAs coupled double quantum dot system, representing a TLS. The two lowest branches are the singlet and triplet states, respectively, and by changing the detuning potential we can tune their splitting. The splitting is also strongly dependent on the interdot distance, which also affects the localization of the two states}
\label{fig:many_body}
\end{figure}

The many-body Hamiltonian is then assembled and diagonalized, allowing us to extract quantities such as coherent oscillations, Coulomb blockade diagrams, and interaction integrals, which can be analyzed standalone or fed into higher-level quantum dynamics models, based on e.g. master equations, to simulate the qubit operation and decoherence due to coupling to the environment. We can also vary the detuning potential (i.e., the difference in potential depth between the two quantum dots) to see at which values the electrons localize on either dot and form singlet or triplet states (Fig. \ref{fig:many_body}). The transitions between these states define the operations of read-out or manipulation of the qubit. \\ 

% ==================
% # IV. CONCLUSION #
% ==================

\section{CONCLUSION}
We have illustrated how to employ atomistic modeling as a tool to guide qubit development by screening novel materials for desired properties, and to obtain a deeper understanding of the intrinsic physical behavior of these complex materials. We do so by using a combination of discrete models such as forcefields (in some cases machine-learned), tight-binding, and advanced first-principles method coupled with Green’s functions to model the structural, thermal, electronic and other properties. A key benefit of atomistic modeling is that the simulations can distinguish the influence of effects which in experiments are difficult to observe directly or isolate from each other. For example, in an experiment you may not know if the destruction of a topological edge state is due to the presence of impurities or the result of intrinsic, spontaneous breaking of time-reversal symmetry at a particular edge, as discussed in one example here. Another utilization of in silico experiments is to easily and cost-efficiently screen many more potential material candidates than is realistic to synthesize and measure in a lab. \\ 

Specifically in the field of superconductivity, the QuantumATK tool enables precise modeling of superconducting tunnel junctions, which are critical for quantum computing and high-sensitivity sensors. By simulating the atomic-scale variability in junctions, researchers can optimize these devices for improved performance and reliability. As quantum technologies continue to advance, the ability to accurately model and predict material behaviors at the atomic level becomes increasingly important to drive innovations in superconductivity, and beyond, ultimately contributing to the development of more efficient, scalable, and reliable quantum devices. \\

% ==============
% # REFERENCES #
% ==============

\section{ABOUT SYNOPSYS}

Founded in 1986 in North Carolina, USA, Synopsys is now among the 12 largest software companies in the world
and a world leader in the areas of Electronic Design Automation (EDA) and Technology Computer Aided Design (TCAD). Headquartered in Sunnyvale, California, Synopsys employs over 30,000 engineering and support staff around the world.

\end{document}